\documentstyle[12pt]{article}
\begin{document}
\begin{center}
{\bf ELEMENTARY ATOM INTERACTION WITH MATTER\footnote{Talk given at 
the International Workshop {\it Hadronic Atoms and Positronium in
the Standard Model}, Dubna, May 26-31, 1998.}\\}
\vspace*{1cm}
STANIS\L AW MR\' OWCZY\' NSKI\footnote{e-mail: mrow@fuw.edu.pl}\\[2mm]
{\it So\l tan Institute for Nuclear Studies\\
ul. Ho\. za 69, PL - 00-681 Warsaw, Poland \\
and Institute of Physics, Pedagogical University\\
ul. Konopnickiej 15, PL - 25-509 Kielce, Poland\\}

\end{center}
\vspace*{.5cm}
\begin{abstract}

{\small The calculations of the elementary atom (the Coulomb bound state 
of elementary particles) interaction with the atom of matter, which are 
performed in the Born approximation, are reviewed. We first discuss the 
nonrelativistic approach and then its relativistic generalization. The 
cross section of the elementary atom excitation and ionization as well 
as the total cross section are considered. A specific selection rule, 
which applies for the atom formed as positronium by particle-antiparticle 
pair, is analyzed.}

\end{abstract}

\vspace*{1cm}

The aim of my lecture is to discuss how the elementary atom, which 
is the Coulomb bound states of elementary particles, interacts 
with the atom of matter. The problem is interesting by itself but
the main motivation comes from the experimental studies of 
$A_{\pi \mu}$ \cite{Coo76}, $A_{2e}$ \cite{Ale84,Afa89} and 
$A_{2\pi}$ \cite{Afa93,Afa94,Ade95}. $A_{ab}$ denotes the atom of the 
positively charged particle $a$ and the negative particle $b$. The atom 
built of the particle $a$ and its antiparticle $\bar a$ is written as 
$A_{2a}$. It should be said that I am going to review my calculations 
which are over ten years old. Let me note however that the results 
have never been presented at a conference. When the series of 
my papers \cite{Mro85,Mro86,Mro87,Den87,Bog88} was published, very a few 
people were interested in the topic -- it was apparently too early. 
Now, after a decade it is somewhat too late since there have appeared 
studies beyond the Born approximation which I used. Nevertheless, the 
results obtained at the Born level remain a reference point for more 
elaborated approaches and are still of interest. Since my calculations 
were all published long ago, I will present only the results and stress 
the most important points. I would like to start however with a few 
personal recollections. 

It was about 15 years ago when I got to know Leonid Nemenov. We met not 
on a professional ground but due to our children who played together. 
We were often walking along the Volga river and talking about politics 
and physics. Leonid introduced me the physics of elementary atoms. He also 
suggested me to perform systematic calculations of the elementary atom 
interactions with matter. At that time I worked as a Ph. D. student in 
Dubna. I was experimentalist involved in the relativistic heavy-ion 
physics. I had already published a few phenomenological papers but wanted 
to be a real theorist and dreamt to work with the field theory. The 
elementary atom calculations seemed to be a good practical course of QED. 
At the beginning I reviewed the existing literature. It was not 
difficult because by that time there had been published all together only 
four papers \cite{Dul77,Pra79,Kot80,Dul83}, where the cross sections of
the elementary atom interaction with matter atom were calculated. 
I realized soon that the two earlier publications \cite{Dul77,Pra79} were 
erroneous while the latter ones \cite{Kot80,Dul83} rather 
unsatisfactory \cite{Mro86}. So, the systematic calculations were indeed 
needed. I started with the nonrelativistic calculations and then worked 
out the relativistic generalization. Let me briefly present the results.

Within the nonrelativistic Born approximation one easily finds the 
cross section of the elementary atom excitation from the state $(nlm)$
to $(n'l'm')$ as
\begin{eqnarray}\label{exci-el}
d\sigma^{n'l'm'}_{nlm} = {1 \over 2\pi {\bf v}^2}\:\vert U({\bf q}) \vert^2
\vert F^{n'l'm'}_{nlm}(\eta {\bf q}) - 
F^{n'l'm'}_{nlm}(\zeta {\bf q}) \vert^2 q dq \;,
\end{eqnarray}
where ${\bf v}$ is the atom relative velocity; ${\bf q}$ is the momentum 
transfer and $q \equiv \vert {\bf q} \vert$; $U({\bf q})$ represents the 
potential generated by the atom of matter; $\zeta \equiv m_1/M$ and 
$\eta \equiv - m_2/M$ with $m_1$, $m_2$ and $M$ being the masses of, 
respectively, the atom components and the atom itself. Due to the smallness
of the atom binding energy $\zeta - \eta = 1$. The transition form factor
is defined as
\begin{eqnarray*}
F^{n'l'm'}_{nlm}({\bf q}) = \int d^3r \: e^{i{\bf q r}}
\phi_{n'l'm'}^*({\bf r})
\phi_{nlm}({\bf r}) \;,
\end{eqnarray*}
where $\phi_{nlm}({\bf r})$ is the wave function of the elementary
atom internal motion. 

Except the paper \cite{Bog88}, where the elementary atom interaction with 
hydrogen was studied, I treated the atom of matter as a structureless 
source of the electromagnetic potential of the Yukava or Thomas-Fermi-Molier 
form. Therefore, the so-called incoherent interactions which lead to the 
target atom excitations were not taken into account. The target recoil 
was neglected as well. Such an approximation is justified for a sufficiently 
heavy matter atom \cite{Mro86}. The role of the incoherent interactions was
studied by other authors \cite{Pak85,Kup89}. 

When the atom is composed, as positronium, of particle and antiparticle,
the cross section (\ref{exci-el}) is nonzero if the atom state parities at 
the initial and final states differ from each other i.e. if
\begin{eqnarray}\label{selection}
(-1)^l = -(-1)^{l'} \;.
\end{eqnarray}
This happens because of the relation
\begin{eqnarray*}
F^{n'l'm'}_{nlm}(-{\bf q}) = (-1)^{l-l'} F^{n'l'm'}_{nlm}({\bf q})\;,
\end{eqnarray*}
which follows from the parity properties of the hydrogen-like atom
wave functions. When the masses of the atom constituents are close
to each other, as in the case of $A_{\pi \mu}$, the transitions, which
break the selection rule (\ref{selection}), are strongly damped. 

The ionization cross section is analogous to the excitation one 
(\ref{exci-el}) and reads
\begin{eqnarray}\label{ion}
d\sigma^{\bf k}_{nlm} = {1 \over (2\pi)^4 {\bf v}^2}\:
\vert U({\bf q}) \vert^2 \vert F^{\bf k}_{nlm}(\eta {\bf q}) - 
F^{\bf k}_{nlm}(\zeta {\bf q}) \vert^2 q dq \: d^3{\bf k} \;,
\end{eqnarray}
with the transition from-factor defined as
\begin{eqnarray}\label{form-ion}
F^{\bf k}_{nlm}({\bf q}) = \int d^3r \: e^{i{\bf q r}}
\phi_{\bf k}^*({\bf r}) \phi_{nlm}({\bf r}) \;,
\end{eqnarray}
where $\phi_{\bf k}({\bf r})$ is the wave function of the ionized 
elementary atom with ${\bf k}$ being the relative momentum of the atom 
components. 

The minimal and maximum momentum transfer are determined by the reaction
kinematics. However, one can take $q_{min}=0$ and $q_{max} = \infty$
as long as the elementary atom is sufficiently energetic in the initial 
state \cite{Mro86,Mro87}. When the minimal and maximal values of $q$ are
assumed to be the same for all final states, the total cross section
can be easily computed due to the sum rule
\begin{eqnarray}\label{sum}
\sum_f \vert F^f_{nlm}(\eta {\bf q}) - 
F^f_{nlm}(\zeta {\bf q}) \vert^2 = 2 - 2 F^{nlm}_{nlm}({\bf q}) \;,
\end{eqnarray}
where the summation runs over the complete set of the quantum states.
Then, the total cross section reads
\begin{eqnarray}\label{total}
\sigma^{tot}_{nlm} = {1 \over \pi {\bf v}^2} \int_0^{\infty} dq \,q\,
\vert U({\bf q}) \vert^2  \big[ 1 - F^{nlm}_{nlm}({\bf q}) \big] \;.
\end{eqnarray}

When the ionization cross section is computed, one is tempted to
substitute the plane-wave function into the form factor (\ref{form-ion}).
In this case however, the cross section which is obtained by integration
of the expression (\ref{ion}) equals the total cross section not the 
ionization one. The point is that the integration over the plane wave 
momentum corresponds to the summation over the complete set of quantum 
states as in eq. (\ref{sum}). Since using the exact scattering Coulomb 
wave function is rather cumbersome, Pak and Tarasov \cite{Pak85} computed 
the ionization cross section subtracting the elastic and excitation 
contributions from the total cross section.

The relativistic generalization of the results presented above is far 
not straightforward due to the well known difficulties of the relativistic 
treatment of the bound states. In particular, the bound state internal 
motion cannot be factorized from the motion of the center of mass. 
However, in the case of the elementary atoms, which are loosely bounded, 
the difficulties can be circumvented to a large extent when the interaction 
is studied in the reference frame where the elementary atom is initially 
at rest. Then, the atom internal motion in the initial as well as in the 
final state is basically nonrelativistic. The point is that the characteristic
momentum transfer to the atom is of order of the inverse Bohr 
radius \cite{Mro87}.

Within the relativistic approach it is desirable to distinguish between 
the spin$-\!{1 \over 2}$ and spinless atom components. Then, we have 
spin$-\!{1 \over 2}\!-\!{1 \over 2}$ atoms such as $A_{2e}$, $A_{e\mu}$, 
the spin$-\!0\!-\!{1 \over 2}$ as $A_{\pi e}$, $A_{K\mu}$ and finally the 
spin$-\!0\!-\!0$ as $A_{2\pi}$ or $A_{\pi K}$. In the case of the 
spin--0--0 atoms, one finds the excitation cross section as 
\begin{eqnarray}\label{exci}
d\sigma^{n'l'm'}_{nlm} = {Z^2 e^4 \over (2\pi)^2 {\bf v}^2}\:
\big| \Delta(Q) \big|^2 \:
\Big| \Big( 1 &+& {{\bf q v} \over 2M} \Big)
\Big[ F^{n'l'm'}_{nlm}(\eta {\bf q}) 
- F^{n'l'm'}_{nlm}(\zeta {\bf q}) \Big] \\[2mm] \nonumber
&-&{{\bf v} \over 2M}
\Big[ {1 \over \zeta} {\bf G}^{n'l'm'}_{nlm}(\eta {\bf q}) 
- {1 \over \eta}{\bf G}^{n'l'm'}_{nlm}(\zeta {\bf q}) \Big]
\Big|^2 q dq \, d\phi \;,
\end{eqnarray}
where $Ze$ is the electric charge of the matter atom nucleus and 
$\Delta(q)$ is the photon propagator in the Lorentz gauge which takes 
into account the effect of screening; $Q$ is the four-momentum transfer 
and $\phi$ is the azimuthal angle of ${\bf v}$ with respect to the 
quantization axis; ${\bf G}^{n'l'm'}_{nlm}$ is the magnetic form factor 
defined as
\begin{eqnarray}\label{form-mag}
{\bf G}^{n'l'm'}_{nlm}({\bf q})
= i\int d^3r \: e^{i{\bf q r}}
\phi_{n'l'm'}^*({\bf r})\, \nabla
\phi_{nlm}({\bf r}) \;.
\end{eqnarray}
The main difference between the relativistic formula (\ref{exci}) and
its nonrelativistic counterpart (\ref{exci-el}) is the appearance of 
the magnetic contribution which can be sizeable.

The formula (\ref{exci}) also holds for the spin$-\!0\!-\!{1 \over 2}$
and spin$-\!{1 \over 2}\!-\!{1 \over 2}$ atoms when the interaction
does not change the atomic spin or spin projection. The cross section
of the spin flip process $(s \rightarrow -s)$ of the 
spin$-\!0\!-\!{1 \over 2}$ atom is
\begin{eqnarray}\label{flip}
d\sigma^{n'l'm'-s}_{nlms} = {Z^2 e^4 \over 8\pi {\bf v}^2}\:
\big| \Delta(Q) \big|^2 \:
\Big| F^{n'l'm'}_{nlm}(\zeta {\bf q}) 
\Big|^2 \,{q^3 \sin^2\alpha \over m_2^2} \,dq  \;,
\end{eqnarray}
where $\alpha$ is the angle between the quantization axis, which
coincides with the vector ${\bf q}$, and ${\bf v}$. Finally, one
finds the cross section of the spin$-\!{1 \over 2}\!-\!{1 \over 2}$ atom
interaction with the change of the atomic spin $(\sigma)$ and/or spin
projection $(\sigma_3)$. The cross section vanishes if (1) 
$\sigma \not= \sigma'$ and $\sigma_3 = \sigma_3'$, (2) 
$|\sigma_3' - \sigma_3|> 1$. In all other cases
\begin{eqnarray}\label{spin-change}
d\sigma^{n'l'm'\sigma'\sigma_3'}_{nlm\sigma'\sigma_3'} 
= {Z^2 e^4 \over 16\pi {\bf v}^2}\:
\big| \Delta(Q) \big|^2\,
\Big|{1 \over \zeta} F^{n'l'm'}_{nlm}(\eta {\bf q}) + 
(-1)^{\sigma'-\sigma} {1 \over \eta}F^{n'l'm'}_{nlm}(\zeta {\bf q}) 
\Big|^2 \,{q^3 \sin^2\alpha \over M^2}\, dq  \;.
\end{eqnarray}
This formula can be used to compute, in particular, the transition
from ortho- to para-positronium. The total cross sections can be found 
from eqs. (\ref{exci}), (\ref{flip}), and (\ref{spin-change}) by means 
of the sum rules analogous to (\ref{sum}).

In the relativistic approach the selection rule (\ref{selection}),
which applies for the atoms composed by particle-antiparticle pairs,
gets a more general form: The transition is allowed if the charge 
parity of the atom changes in the course of interaction i.e.
$$
(-1)^{l+\sigma} = - (-1)^{l'+\sigma'}\;.
$$ 
The reason is the following. The atom of particle and antiparticle 
is an eigenstate of the charge parity operator with the eigenvalue 
$(-1)^{l+\sigma}$. The photon is also the parity eigenstate with 
the eigenvalue $-1$. Since the Born approximation corresponds to the 
one-photon exchange the charge parity conservation leads to the selection 
rule of interest.

The numerical results of the cross sections were collected in 
the papers \cite{Mro86,Den87}. Interaction of eight elementary atoms
($A_{2e}$, $A_{e\mu}$, $A_{e\pi}$, $A_{2\mu}$, $A_{\mu\pi}$, $A_{2\pi}$,
$A_{\pi,K}$, and $A_{2K}$) with five targets (C, Al, Cu, Ag, Pb)
was studied. The excitation and total cross sections were calculated. 
The electric, magnetic\footnote{Drs. Afanasyev and Tarasov have recently
informed me that my results concerning the magnetic contributions to the 
cross sections are numerically underestimated for the heaviest elementary 
atoms.}, spin and para-ortho transitions were analysed separately. 

The calculations of the elementary atom cross sections presented here 
were improved by several authors. I have already mentioned the 
works \cite{Pak85,Kup89}, where the role of the incoherent interactions 
was studied. In the papers \cite{Tar91,Vos97} the multiphoton exchanges 
within the eikonal approach were taken into account. Afanasyev and 
Tarasov \cite{Afa96} discussed the interaction of the excited atoms.
Finally, I should mention very valuable calculations presented at this 
Workshop by Afanasyev, Cugnon, Tarasov and Trautmann. 


\end{document}